\documentclass[12pt]{article}
\usepackage[utf8]{inputenc}
\usepackage{graphicx}
\usepackage{grffile}
\usepackage{lscape}
\usepackage{url}
\usepackage{marvosym}
\usepackage{abstract}
\newcommand{\footremember}[2]{%
    \footnote{#2}
    \newcounter{#1}
    \setcounter{#1}{\value{footnote}}%
}
\newcommand{\footrecall}[1]{%
    \footnotemark[\value{#1}]%
} 

\usepackage{hyperref}

\providecommand{\keywords}[1]
{
  \small	
  \textbf{\textit{Keywords---}} #1
}

\usepackage{geometry}
\hypersetup{
    colorlinks=true, 
    linktoc=all,     
    linkcolor=blue,  
}
\title{Territorial differences in the spread of COVID-19 in European regions and US counties}

\author{Fabrizio Natale\footnote{Corresponding author, Email: \href{mailto:stefano.iacus@ec.europa.eu}{fabrizio.natale@ec.europa.eu} } \footremember{aldo}{European Commission, 
              Joint Research Centre,
              Via Enrico Fermi 2749, 21027 Ispra (VA), Italy}\and 
Stefano Maria Iacus\footrecall{aldo} \and
Alessandra Conte  \footrecall{aldo} \and Spyridon Spyratos\footrecall{aldo}   \and Francesco Sermi\footrecall{aldo}  
}

\begin{document}

\maketitle

\begin{abstract}
This article explores the territorial differences in the onset and spread of COVID-19 and the excess mortality associated with the pandemic, across the European NUTS3 regions and US counties. Both in Europe and in the US, the pandemic arrived earlier and recorded higher Rt values in urban regions than in intermediate and rural ones. A similar gap is also found in the data on excess mortality. In the weeks during the first phase of the pandemic, urban regions in EU countries experienced excess mortality of up to 68pp more than rural ones. We show that, during the initial days of the pandemic, territorial differences in Rt by the degree of urbanisation can be largely explained by the level of internal, inbound and outbound mobility. The differences in the spread of COVID-19 by rural-urban typology and the role of mobility are less clear during the second wave. This could be linked to the fact that the infection is widespread across territories, to changes in mobility patterns during the summer period as well as to the different containment measures which reverse the causality between mobility and Rt.    
\end{abstract}


\section*{Introduction}
The current COVID-19 pandemic is creating severe social and economic consequences, with some places experiencing disproportionately high levels of mortality and economic losses. Urban regions, and particularly large cities, have been severely affected by the spread of the pandemic in its early stages. The evolving public discussion on the territorial impact of the pandemic requires a greater understanding of the way the pandemic is affecting regions that are diversely vulnerable and will require different recovery plans. 
Analysis of the role of population density on the virus spread has led to mixed results \cite{stier_covid-19_2020,ribeiro_city_2020,heroy_metropolitan-scale_2020}. In addition to density, particular attention has also been given to some factors related to the urban organisation that would make some places vulnerable to infection in the first phase. In particular, these are the connectivity of cities as hubs of national and international transport systems \cite{Mazzoli2020,gerritse_michiel_cities_2020,chinazzi_effect_2020}, the structure of employment and industry \cite{almagro_determinants_nodate,agnoletti_covid-19_2020}, as well as overcrowded living conditions in some home environments \cite{bayer_intergenerational_2020,belloc_cross-country_2020}.  
Overall, this analysis is aimed at a deeper understanding of the links between COVID-19, urban-rural typologies, territorial conditions, and mobility, which is critical for designing effective public health policy responses. We first explore the heterogeneity of COVID-19 patterns in its onset, spread, and associated excess mortality by comparing the results by the level of urbanisation of European regions and counties in the US. The results show that the pandemic started earlier in urban regions than in intermediate and rural areas. Urban regions had the highest Rt values in both Europe and the US during the first wave, whereas rural counties were more affected than urban counties in the second wave. Analysis of excess mortality, calculated using Eurostat statistics and obtained from the difference between reported fatalities and a baseline model based on historical data between 2011 and 2019, also shows a large gap by urbanisation level during the first wave, with a median excess mortality up to 73\% for urban regions, 18\% for intermediate regions, and 11\% for rural regions.   
In a second phase, we empirically examine the impact of domestic mobility on virus spread. We model population mobility in European regions through a harmonised mobility index derived from mobile phone data. We examine the geographical distribution of mobility changes through regression models for the weeks in the first and second virus waves. Our results show that, on the one hand, higher mobility explains most of the variation in values in the weekly Rt during the first wave, with internal, inbound, and outbound mobility positively affecting Rt. The effect of internal mobility, in particular, is more pronounced than that of the degree of urbanisation, and remains significant even when population and population density are taken into account. On the other hand, the same regression models replicated for the second wave weeks show a negative role of mobility on the local spread of the virus, as well as a higher prevalence of the infection in rural regions compared to large cities during the second wave.  
The paper is organised as follows. The data section describes the data and methods used in the analyses. In the results and discussion section, we present how the COVID-19 pandemic was spread in rural, intermediate, and urban regions during the first and the second wave, and finally, the conclusions are outlined in the final section.

\section*{Data and methods}

In this section we present the data sources and the methods we used to assess the spread of the COVID-19 pandemic in rural, intermediate and urban regions. We calculated the reproductive number (Rt) as an indicator to assess how fast the virus spread across different types of geographical areas. We estimated the excess mortality to monitor in quantitative terms the evolution and impacts of COVID-19 pandemic. We used fully anonymised and regionally aggregated mobility data to get insights about the different regional mobility patters. Finally, we fitted a linear regression model to study the relationship between mobility and Rt during the first and the second wave. 

\subsection*{Rt}
Rt is the main real-time indicator used to assess the evolution of the pandemic, design containment measures and monitor their effectiveness.
A time-dependent reproduction number, Rt, was calculated with the R package R0 following a likelihood-based estimation procedure \cite{wallinga_different_2004} with parameters for generation time distribution reported in \cite{zhanwei_du_serial_2020}. and on the basis of daily data on confirmed COVID-19 cases at regional level downloaded through the 'COVID19' R package. The daily Rt values were aggregated to weekly averages.

\subsection*{Excess mortality}
The baseline for mortality was calculated with Generalised Additive Models fitted independently for each region. These models include a seasonal component to account for the increase in mortality during the winter months linked to influenza outbreaks, and a linear time trend to account for long-term changes in mortality due to demographic dynamics. The excess mortality is measured as difference between the reported data in 2020 and the estimated baseline for all occurrences exceeding the lower or upper 95\% confidence intervals of the estimated baseline. The weekly mortality are from Eurostat (demormweek3) and cover 900 regions in 26 EU MS and the UK with time series spanning from 2001 to 2015.

\subsection*{Mobility}
In this study we used fully anonymised and aggregated mobility data shared with the European Commission (EC) by European Mobile Network Operators (MNOs). These mobility data comply with the `Guidelines on the use of location data and contact tracing tools in the context of the COVID-19 outbreak' by the European Data Protection Board \cite{edpb2020}. The mobility data are in the form of Origin-Destination Matrix (ODM) \cite{ODMs2019, ODMs2020} and they provide valuable insights into mobility patterns across geographical areas. The data has been used to derive mobility insights and build tools to inform better targeted containment measures, in a Mobility Visualisation Platform, available to the Member States  \cite{eu_com}.\\ Given the high variation in the spatial and temporal aggregation across countries and operators, the original ODMs are harmonised at standardised spatial and temporal granularity to the derived Mobility Indicators. For more information about the Mobility Indicators we direct the interested reader to  \cite{SANTAMARIA2020104925}. In this study, we further aggregated the Mobility Indicators at a weekly temporal granularity, and we normalised them in both per capital and absolute terms to enable a better cross-country comparison on the number of recorded movements. This further normalisation was performed by setting for each country and type of mobility (internal, inbound, outbound) the value of one for the NUTS3 regions with the higher mobility over the reference time period February to December 2020, and the value of zero to the NUTS3 regions with the lowest mobility over the same time period.

\subsection*{Regression}
The fitting of the regression models was constrained by the necessity of having regional data on COVID-19 cases for the calculation of Rt and mobility indicators for the same periods. Data on population was obtained from Eurostat (demorpjangrp3 and demord3dens). Overall the regressions are based on around 3500 observations in 654 regions for the first wave, and 10500 observations in 551 regions for the second wave. The regression models include country fixed effects to control for country-specific and time-invariant factors that may influence the spread of the disease. 

\section*{Results and Discussion}

\subsection*{The COVID-19 pandemic started earlier in urban regions} 
\begin{figure}[ht]
\centering
\includegraphics[width=1\linewidth]{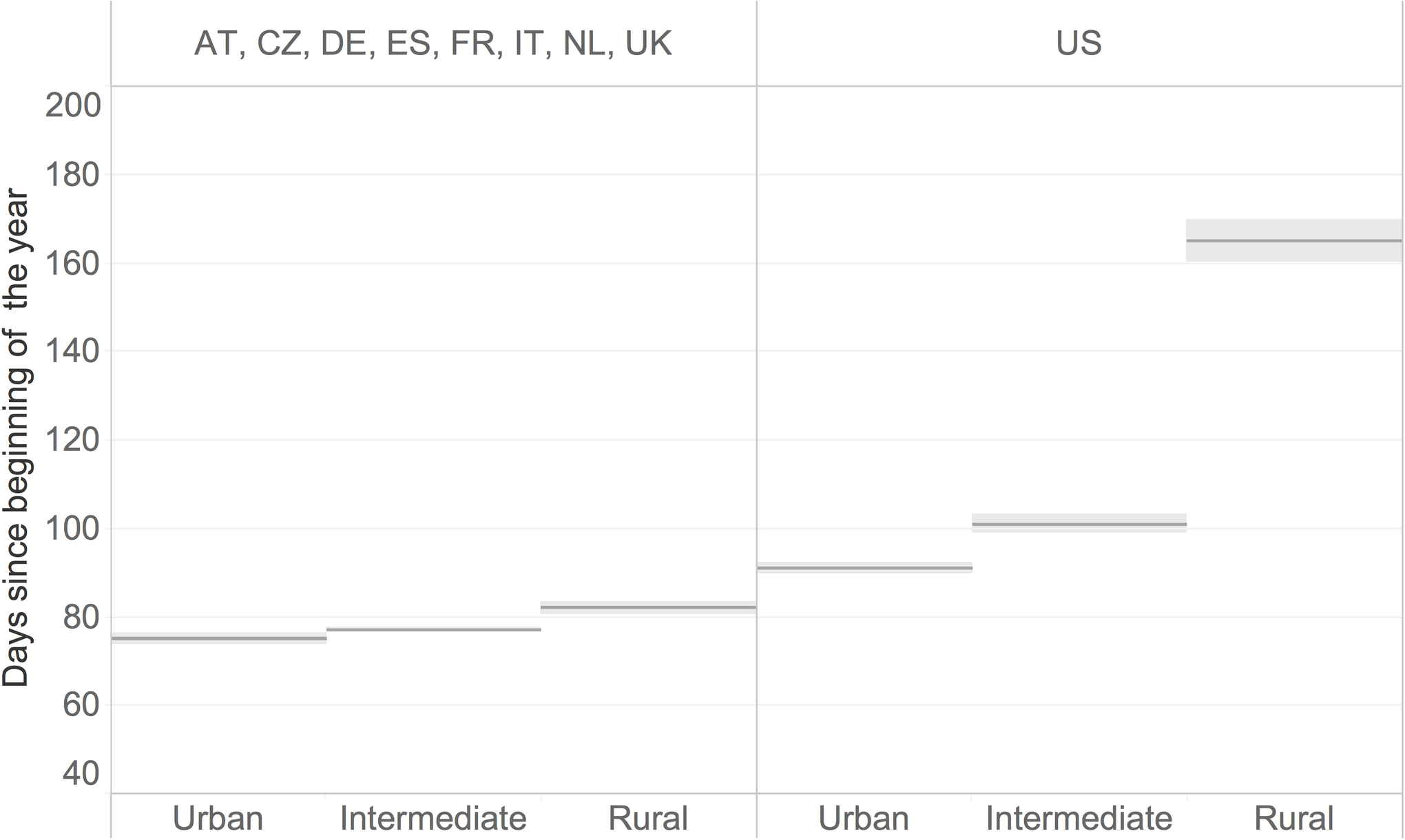}
\caption{Onset of the pandemic across regions by rural-urban typology.}
\label{fig:fig1}
\end{figure}

Figure \ref{fig:fig1} describes the pandemic onset in the NUTS3 regions of some European countries and counties in the US, clustered by the three levels of urbanisation (for the EU we use Eurostat NUTS3 rural-urban typologies and for the US we use Rural-Urban Continuum Codes reduced to 3 classes). We measure the pandemic onset in each region by the number of days between the registration of the first 20 confirmed cases of Coronavirus disease and the beginning of the year 2020. In both Europe and the United States, urban regions are more vulnerable to the pandemic's onset. The pandemic started earlier in most urban regions, while we observe a later onset in intermediate and rural regions in the first wave of the virus.

\subsection*{The infection has spread faster in urban regions during the first wave}

\begin{figure}[htb]
\centering
\includegraphics[height=0.5\textheight]{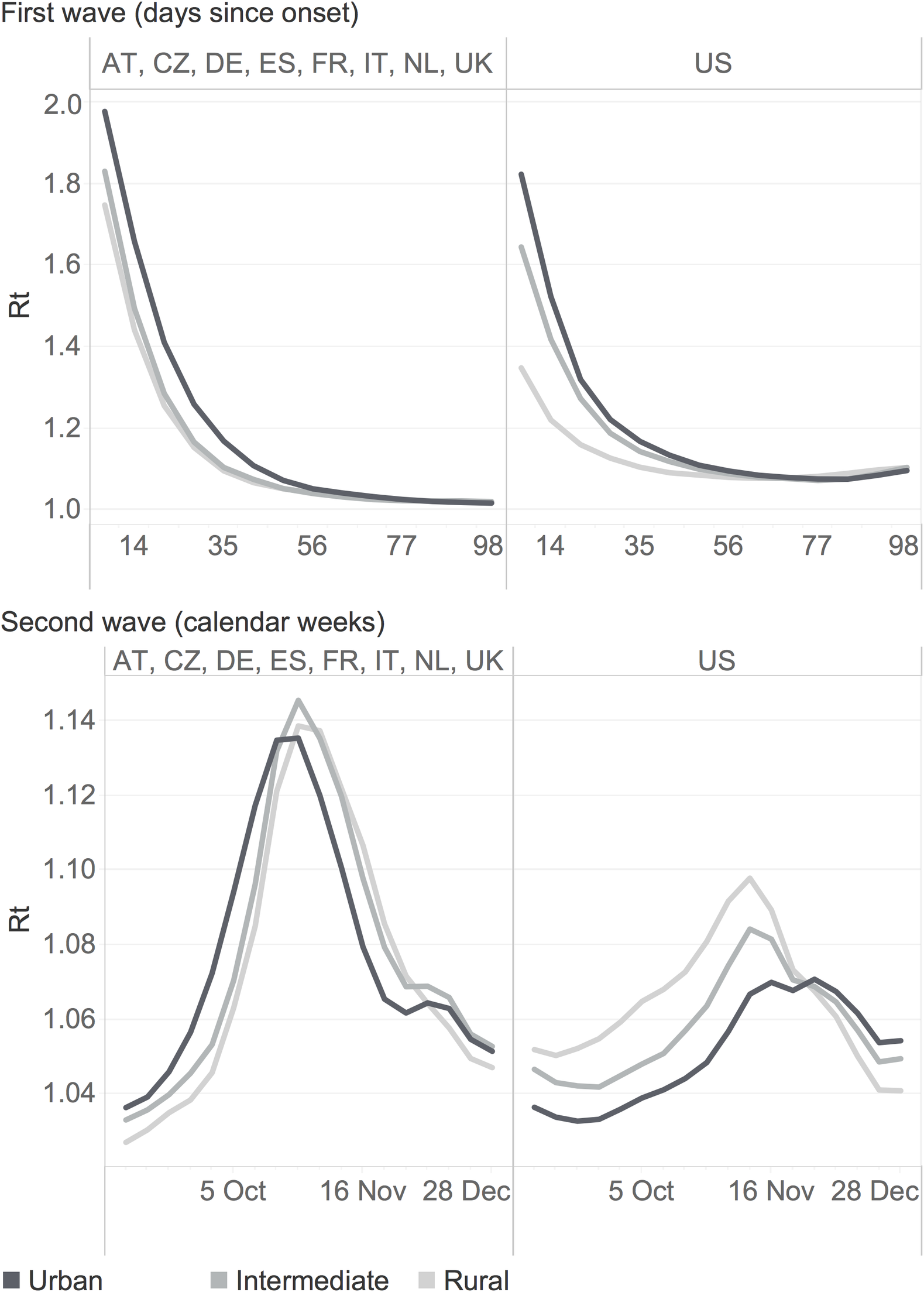}
\caption{Median Rt values in European regions and American counties by rural-urban typology. The upper part presents the daily Rt values from the first recorded case in each region during the first wave. The lower part represents the weekly Rt values during the second wave.}
\label{fig:fig2}
\end{figure}

 Figure \ref{fig:fig2} displays the Rt values for the first and second waves of the pandemic. Rt is calculated from daily confirmed cases in 807 NUTS3 regions in the UK, Netherlands, Germany, Italy, Spain, France, Czech Republic and Austria (left) and in 3100 counties in the US (right). The indicator is averaged across regions grouped by rural-urban type and aggregated by days since the first reported case in each region (upper Figure), and weeks since the start of the second wave of the pandemic (lower Figure).  
Looking first at the upper figure, we observe that urban regions in Europe and the United States recorded higher Rt values than those found in intermediate and rural regions at the start of the pandemic. This indicates that the disease spread faster in urban regions and that containment was more difficult in more densely populated areas. Approximately 56 days after the start of the pandemic, we find a general decline in the Rt and a reduction in the differences in Rt between the three groups of regions. At the start of the pandemic, the rural-urban divide in Rt values is more pronounced in the US counties. However, even in this case, the disparity in the pandemic spread by level of urbanisation has narrowed among the three regional groups, with the Rt index close to 1 at the end of the first wave. 
The lower part of Figure \ref{fig:fig2} shows the median Rt values across regions and counties in the weeks following the summer period, when the pandemic began to spread in a second wave of infections. In the European regions, we observe an initially higher Rt in the urban regions and increasing and higher values in the intermediate and rural regions as the second wave progresses. In contrast, in the US, rural and intermediate counties are the most vulnerable to virus spread for most of the weeks during the second wave, with a slight change in trend in the last weeks of the period.

\subsection*{The excess mortality linked to COVID-19 is higher in the European urban regions in the first wave}

\begin{figure}[htb]
\centering
\includegraphics[width=1\linewidth]{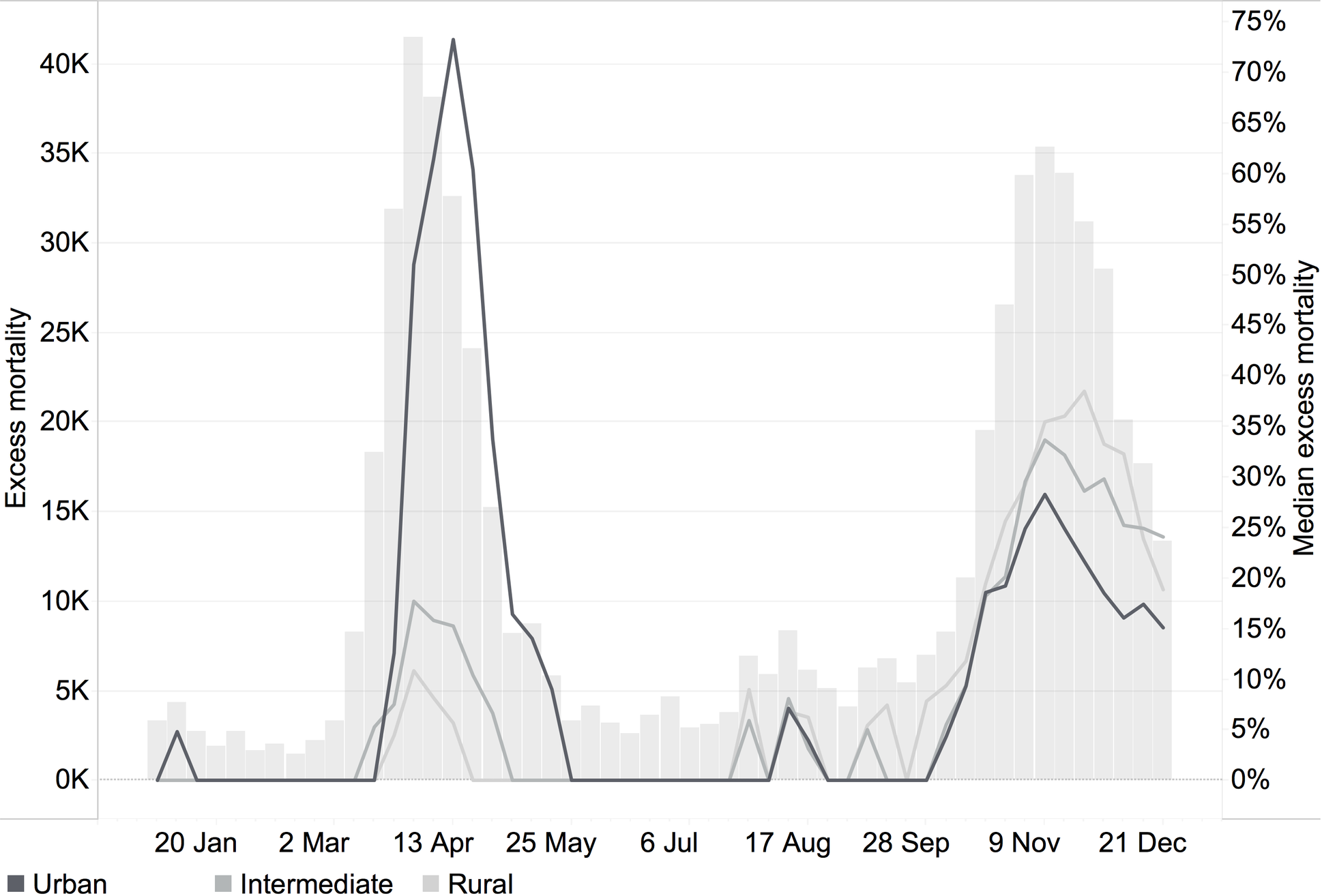}
\caption{Total excess mortality (bars) and median excess mortality by rural-urban typology (lines) on a weekly basis.}
\label{fig:fig3}
\end{figure}

Figure \ref{fig:fig3} shows the trend in the excess mortality for the European regions during the year 2020. The increase in weekly mortality compared to past trends  is used as an indirect measure to monitor the evolution of COVID-19. This indicator has the downside of including fatalities not necessarily linked to COVID-19, such as those caused by the saturation of hospital capacity, but has the advantage of being less influenced by the underestimation of the real infection rate due to asymptomatic cases or differences in testing strategies over time and regions \cite{Bartoszek2020}.
The bars in Figure \ref{fig:fig3} show the weekly total excess mortality calculated from Eurostat statistics for most EU countries and the UK. The excess mortality is obtained from the difference between the reported fatalities and a modelled baseline estimated from historical data between 2011 and 2019. 
The number of weekly fatalities attributable to COVID-19 peaked at the beginning of April, with about 41 400 deaths in excess compared to the baseline.\footnote{This peak represents 21 600 more cases than the excess mortality recorded in the same countries during the second week of January 2017, corresponding to a particularly severe year for the seasonal flu.}
The lines in the figure show the median excess mortality in the NUTS3 regions classified according to their degree of urbanisation. At the peak of the pandemic in third week of April, the median excess mortality in urban regions reached its peak with an excess mortality of 73\%, which was 58 pp higher than in intermediate regions and 68 pp higher than in rural regions in the same week. 
In the second wave of the pandemic, the disparities among regions appear less pronounced. There is also a reverse in the trend of excess mortality, with rural and intermediate regions having higher rates, 38\% and 32\% respectively, than urban regions with an excess mortality rate of 26\%.

\subsection*{Mobility is higher in urban regions}

\begin{figure}[htb]
\centering
\includegraphics[width=1\linewidth]{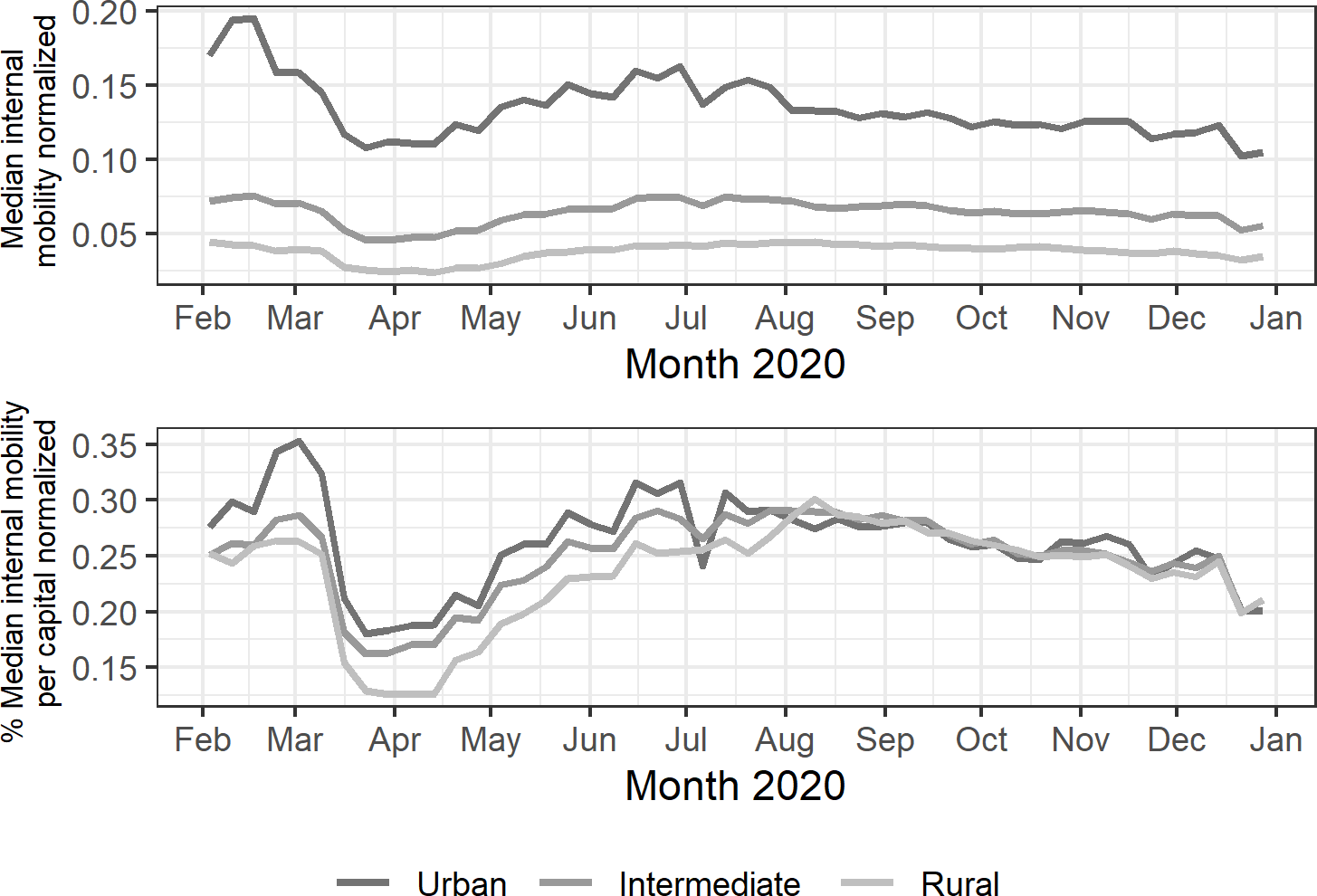}
\caption{Median absolute and per capita internal mobility across regions grouped by rural-urban typology}
\label{fig:fig4}
\end{figure}

\begin{figure}[htb]
\centering
\includegraphics[width=1\linewidth]{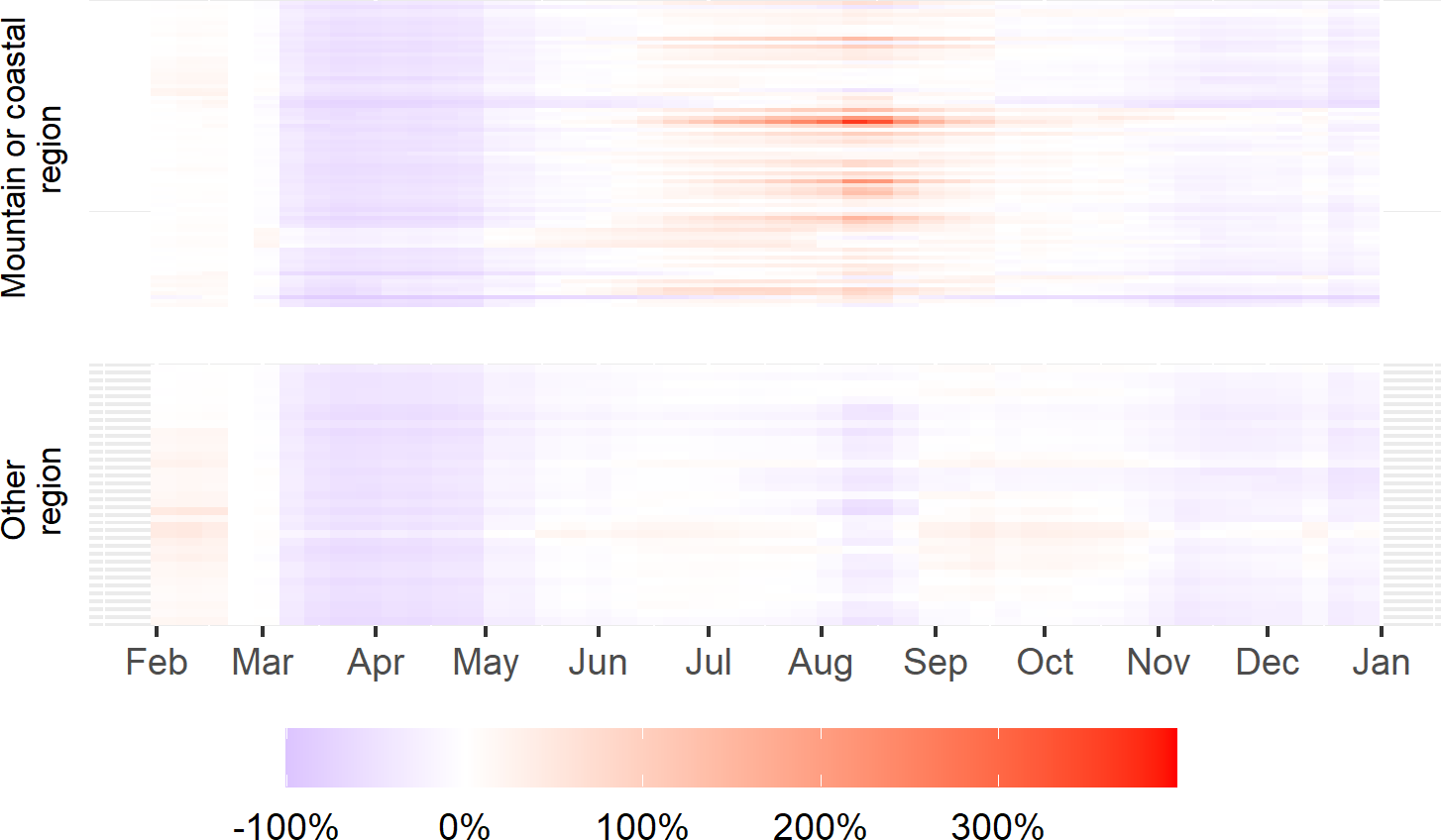}
\caption{Changes in internal mobility for regions in Italy in respect of the level in the last week of February}
\label{fig:fig5}
\end{figure}

One possible explanation for the higher Rt and excess mortality in urban regions is that in cities the infection can spread more rapidly given the higher population density, larger use of public transportation and higher number of social interactions. The intensity of social interaction is reflected in mobility indicators which can be calculated from social media and mobile phones data. 
In fact, the relation between intensity of social contacts, mobility and infection is at the basis of mobility restriction that most governments have put in place to contain the pandemic.
We analyse the patterns of mobility within, from and toward European regions with anonymised and aggregated mobile indicators derived from mobile phone data as described in the Data and methods Section.
Figure \ref{fig:fig4} shows the median patterns of weekly mobility of 1033 NUTS3 regions in 22 EU countries, grouped by rural, intermediate and urban typology, in absolute (upper chart) and per capita terms (lower chart). 
The trends in the two charts in Figure \ref{fig:fig4} reflect the implementation of generalised lockdown until April, the reopening during the summer period and the new restrictions on mobility after summer.
The upper chart in Figure \ref{fig:fig4}, reflects that the mobility in absolute terms in urban areas is higher compared to the intermediate and rural ones, mainly due to their different population size. The lower chart in Figure \ref{fig:fig4} describes that during the first wave, and independently from the implementation of the restriction measures, the level of per capita mobility was higher in urban regions in respect of intermediate and rural ones. During the second wave, the per capita mobility is almost equal across all areas, indicating substantial reduction of mobility in urban and intermediate regions at the beginning of summer and an increase in rural regions.
This shift in mobility patterns is exemplified in Figure \ref{fig:fig5} showing the weekly relative changes in mobility for each Italian region (rows) in respect of the levels recorded during the last week of February (first column). In this case, regions are sorted on the basis of their proximity to the sea or mountains to better appreciate the mobility linked to domestic tourism.
In May, after the lifting of lockdown, all Italian regions recorded an increase of mobility to the levels of February. However, during summer, in coastal and mountain regions mobility increased to higher values than at the beginning of the year. The highest increase was recorded in the second week of August in the renowned region of Olbia in Sardinia (+373\%).
With the re-opening of schools in September, the level of mobility started again to increase uniformly across all regions. 

\subsection*{The higher mobility in urban regions may explain great part of the territorial gaps in Rt during the first wave}
\begin{figure}[htb]
\centering
\includegraphics[height=0.5\textheight]{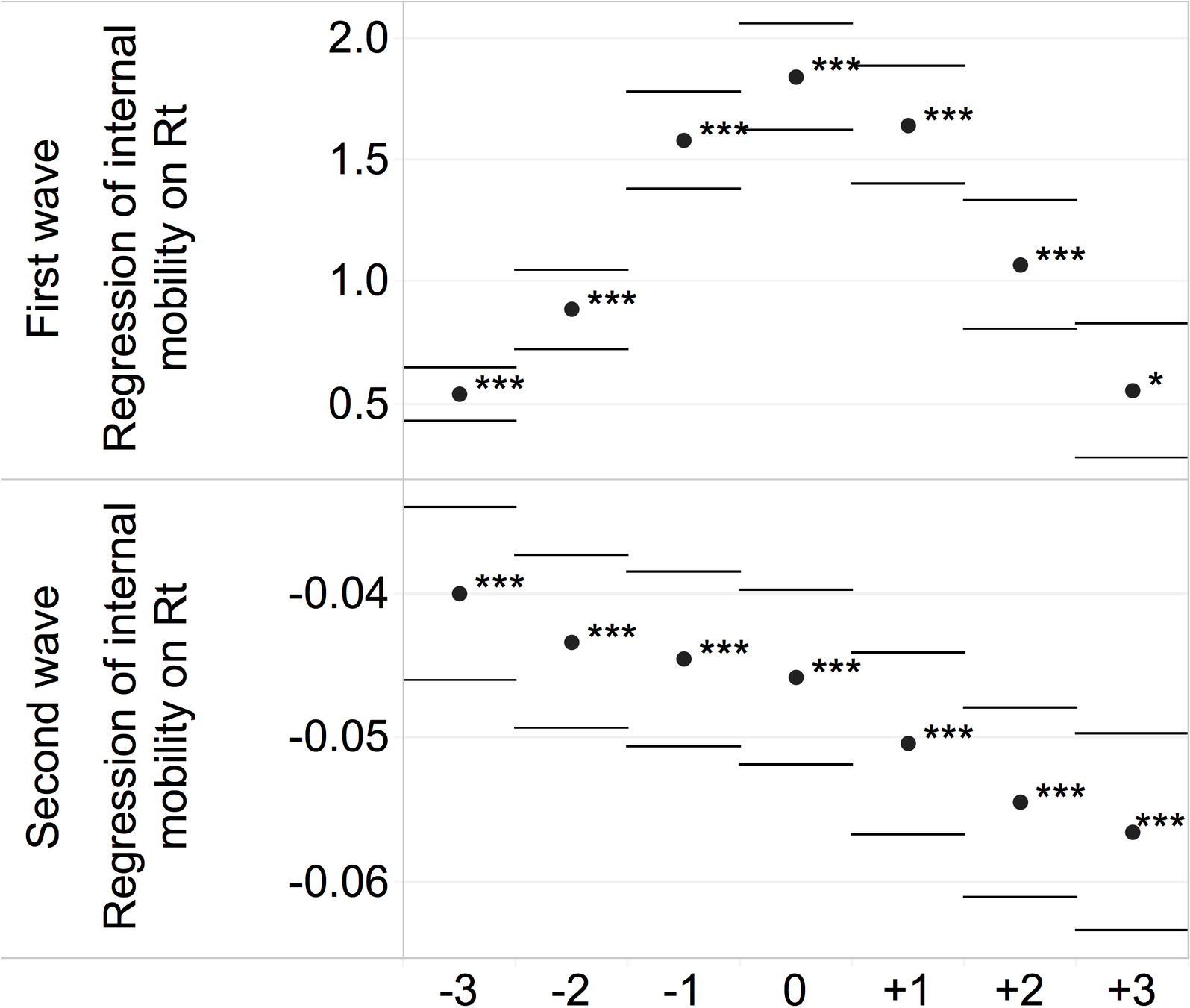}
\caption{Regression coefficients  with shifts of internal mobility of 3 weeks before and after the reference week of Rt.}
\label{fig:fig6}
\end{figure}

To support our intuition about the heterogeneity in the spread of Covid during the first and second waves, we examine the effect of different mobility patterns on the spread of the pandemic through a series of regression models using the Rt values recorded in each European region as the dependent variable. Table \ref{tab:Table1} shows the results of the ordinary least squares regressions on the first wave of infection considering the Rt values in the first 28 days after the start of the pandemic. Table \ref{tab:Table2} presents results for the second wave on the Rt values in the weeks after August and Figure \ref{fig:fig6} provides estimates of the relation between Rt and mobility using different temporal shifts between the two variables.  All specifications include country fixed effects to account for differences in transmission resulting from invariant country characteristics. 
We estimate a significant relationship between the effective reproduction number, Rt, and the levels of urbanisation (Column 1 in Table \ref{tab:Table1}). During the first wave of the pandemic, Rt values are lower in rural and intermediate regions than in the urban regions used as reference. In Columns 2-4 we include the mobility controls separately, i.e. internal, incoming and outgoing mobility, given the correlation between these measures within countries. In all specifications that include controls separately, each mobility indicator is positively correlated with Rt values. Mobility is also analysed using a per capita mobility indicator (Column 5). The positive and significant coefficient of the per capita mobility confirms a pattern of Rt that increases as the internal mobility measured on the total population increases. The demographic controls of the total population and density, presented in Columns 6 and 7, also have a positive effect on Rt in the first wave. Finally, in Columns 8 and 9, we simultaneously estimate the effect of the internal mobility, the level of urbanisation of the regions and the population density. The coefficient of the internal mobility remains significant and positive even when we include the other control variables. However, the variables indicating the level of urbanisation lose their significance while the variables on population size and density change sign. Internal mobility appears to be a critical determinant of the rate of COVID-19 cases during the first wave, positively influencing the spread of the virus possibly through increased social interactions. These results confirm that great part of the territorial characteristics influencing the higher epidemiological risk at the onset of the pandemic in urban regions can be explained by the role of mobility. 
Table \ref{tab:Table2} examines the relationship between Rt and different mobility patterns in the European regions in a similar way to Table 1 but with data for the second wave (from August). The estimates show an inversion of sign in respect the first wave with a positive association between the virus spread and the rural and intermediate regions compared to large cities. These results may reflect a behavioural response as well as more severe containment measures in the most severely affected areas. In the second wave, different mobility patterns are associated with decreased Rt values, as presented in Column 2-5. The results thus indicate that the relationship between mobility and the regional virus transmission has changed over time and that shifts in mobility are used to control the epidemic. However, these changes were not sufficient to prevent a second wave of infection in most of the regions. The population demographic variable is significant and negative on the virus spread. The models that simultaneously estimate the effect of internal mobility and different regional characteristics confirm a negative relationship between this mobility pattern and the virus transmission, as well as a higher prevalence of the infection in rural regions compared to large cities during the second wave. When interpreting the relationship between the virus transmission and the local mobility during the second wave, it is important to consider that mobility changes may respond endogenously to Rt values, as mobility may be related to either individual or government-led changes in preventive behaviour. These estimates may therefore be underestimating the link between mobility and infections during the second wave and a further investigation through an instrumental variables strategy for our specifications would be preferable to address these endogeneity concerns.      
The Figure \ref{fig:fig6} show the regression coefficients  with shifts of internal mobility of 3 weeks before and after the reference week of Rt. During the first wave the relation of mobility on Rt is positive and peaks during the same week. During the second wave the relation is negative and constantly decreasing. The fact that the relation during the first wave is becoming clearer towards the reference week is indicating that mobility is having an effect on Rt. On the contrary during the second wave there is an inversion of causality and mobility rather than influencing seem to be reacting to changes in Rt moving in opposite directions. Intuitively this is in line with the consideration that during the advanced stages of the pandemic  mobility is highly conditioned by restriction measures and lockdowns which are put in place in correspondence with increases in Rt.

\begin{landscape}

\begin{table*}
\centering 
  \caption{Regression on Rt during the first wave (28 days since onset)} 
  \label{tab:Table1} 

\begin{tabular}{@{\extracolsep{5pt}}lccccccccc} 
\\[-1.8ex]\hline 
\hline \\[-1.8ex] 
 & \multicolumn{9}{c}{Rt first wave} \\ 
\cline{2-10} 
\\[-1.8ex] & (1) & (2) & (3) & (4) & (5) & (6) & (7) & (8) & (9)\\ 
\hline \\[-1.8ex] 
 Intermediate & $-$0.12$^{**}$ &  &  &  &  &  &  & 0.05 & $-$0.04 \\ 
  & (0.05) &  &  &  &  &  &  & (0.05) & (0.06) \\ 
  Rural & $-$0.16$^{***}$ &  &  &  &  &  &  & 0.07 & $-$0.06 \\ 
  & (0.06) &  &  &  &  &  &  & (0.06) & (0.07) \\ 
  Internal &  & 1.84$^{***}$ &  &  &  &  &  & 1.99$^{***}$ & 4.54$^{***}$ \\ 
  &  & (0.22) &  &  &  &  &  & (0.25) & (0.40) \\ 
  Inbound &  &  & 2.20$^{***}$ &  &  &  &  &  &  \\ 
  &  &  & (0.17) &  &  &  &  &  &  \\ 
  Outbound &  &  &  & 2.42$^{***}$ &  &  &  &  &  \\ 
  &  &  &  & (0.21) &  &  &  &  &  \\ 
  Internal pca &  &  &  &  & 0.27$^{***}$ &  &  &  &  \\ 
  &  &  &  &  & (0.02) &  &  &  &  \\ 
  Population &  &  &  &  &  & 0.05$^{**}$ &  &  & $-$0.27$^{***}$ \\ 
  &  &  &  &  &  & (0.02) &  &  & (0.03) \\ 
  Population density &  &  &  &  &  &  & 0.04$^{***}$ &  & $-$0.05$^{*}$ \\ 
  &  &  &  &  &  &  & (0.02) &  & (0.02) \\ 
 \hline \\[-1.8ex] 
Observations & 3,572 & 3,572 & 3,570 & 3,571 & 3,572 & 3,572 & 4,550 & 3,572 & 3,572 \\ 
R$^{2}$ & 0.02 & 0.03 & 0.06 & 0.05 & 0.07 & 0.02 & 0.01 & 0.03 & 0.05 \\ 
Adjusted R$^{2}$ & 0.01 & 0.03 & 0.05 & 0.05 & 0.07 & 0.01 & 0.01 & 0.03 & 0.05 \\ 
\hline 
\hline \\[-1.8ex] 
\textit{Note:}  & \multicolumn{9}{r}{$^{*}$p$<$0.1; $^{**}$p$<$0.05; $^{***}$p$<$0.01} \\ 
\end{tabular} 
\end{table*}

\begin{table*}
\centering
\caption{Regression on Rt during the second wave (after August)} 
  \label{tab:Table2} 
\begin{tabular}{@{\extracolsep{5pt}}lccccccccc} 
\\[-1.8ex]\hline 
\hline \\[-1.8ex] 
 & \multicolumn{9}{c}{Rt second wave} \\ 
\cline{2-10} 
\\[-1.8ex] & (1) & (2) & (3) & (4) & (5) & (6) & (7) & (8) & (9)\\ 
\hline \\[-1.8ex] 
 Intermediate & 0.004$^{***}$ &  &  &  &  &  &  & 0.001 & 0.002 \\ 
  & (0.001) &  &  &  &  &  &  & (0.002) & (0.002) \\ 
  Rural & 0.01$^{***}$ &  &  &  &  &  &  & 0.01$^{***}$ & 0.01$^{***}$ \\ 
  & (0.002) &  &  &  &  &  &  & (0.002) & (0.002) \\ 
  Internal &  & $-$0.05$^{***}$ &  &  &  &  &  & $-$0.03$^{***}$ & $-$0.07$^{***}$ \\ 
  &  & (0.01) &  &  &  &  &  & (0.01) & (0.01) \\ 
  Inbound &  &  & $-$0.04$^{***}$ &  &  &  &  &  &  \\ 
  &  &  & (0.005) &  &  &  &  &  &  \\ 
  Outbound &  &  &  & $-$0.04$^{***}$ &  &  &  &  &  \\ 
  &  &  &  & (0.01) &  &  &  &  &  \\ 
  Internal pca &  &  &  &  & $-$0.004$^{***}$ &  &  &  &  \\ 
  &  &  &  &  & (0.001) &  &  &  &  \\ 
  Population &  &  &  &  &  & $-$0.003$^{***}$ &  &  & 0.004$^{***}$ \\ 
  &  &  &  &  &  & (0.001) &  &  & (0.001) \\ 
  Population density &  &  &  &  &  &  & $-$0.0004 &  & 0.0002 \\ 
  &  &  &  &  &  &  & (0.0005) &  & (0.001) \\ 
 \hline \\[-1.8ex] 
Observations & 10,542 & 10,542 & 10,534 & 10,538 & 10,542 & 10,542 & 16,204 & 10,542 & 10,542 \\ 
R$^{2}$ & 0.04 & 0.04 & 0.04 & 0.04 & 0.04 & 0.03 & 0.05 & 0.04 & 0.04 \\ 
Adjusted R$^{2}$ & 0.04 & 0.04 & 0.04 & 0.04 & 0.04 & 0.03 & 0.05 & 0.04 & 0.04 \\ 
\hline 
\hline \\[-1.8ex] 
\textit{Note:}  & \multicolumn{9}{r}{$^{*}$p$<$0.1; $^{**}$p$<$0.05; $^{***}$p$<$0.01} \\ 
\end{tabular} 
\end{table*} 

\end{landscape}

\section*{Conclusion}

In this article we analysed the territorial differences in the onset and spread of COVID-19 and the associated excess mortality, across the European NUTS3 regions and US counties during the first and second COVID-19 wave. During the first wave, the COVID-19 pandemic arrived earlier, recorded higher Rt values and had a higher impact in terms of excess mortality in urban regions compared to the intermediate and the rural ones. In the first wave, mobility influenced the spread of COVID-19, since the higher mobility of urban regions is explaining entirely the differences between the three groups of regions. The fact that these effects are more difficult to recognise in later stages of the pandemic can be tentatively explained by the widespread of the infection, the implementation of restriction measures which invert the causality between mobility and Rt, often applied on a territorial basis, and the more complex mobility patterns experienced during the summer period.

Our findings are in line with previous studies identifying the role of mobility on virus spread in the early stages of the pandemic. To our knowledge, our research is unique in providing a broad geographic coverage and a high level of geographical detail, and in examining the role of regional mobility for the spread of COVID-19 through a unique data set on regional mobility derived from mobile phone data. In terms of policy implication, our research contributes to a better understanding of territorial characteristics of the spread of COVID-19, which is critical for designing effective public health policy responses, often decided at regional level.

\section*{Acknowledgments}
The authors acknowledge the support of European MNOs (among which 3 Group - part of CK Hutchison, A1 Telekom Austria Group, Altice Portugal, Deutsche Telekom, Orange, Proximus, TIM Telecom Italia, Tele2, Telefonica, Telenor, Telia Company and Vodafone) in providing access to aggregate and anonymised data. The authors would also like to acknowledge the GSMA\footnote{GSMA is the GSM Association of Mobile Network Operators.}, colleagues from Eurostat\footnote{Eurostat is the Statistical Office of the European Union.} and ECDC\footnote{ECDC: European Centre for Disease Prevention and Control. An agency of the European Union.}  for their input in drafting the data request. \\ Finally, the authors would also like to acknowledge the support from JRC colleagues, and in particular the E3 Unit, for setting up a secure environment to host and process of the data provided by MNOs, as well as the E6 Unit (the ``Dynamic Data Hub team'') for their valuable support in setting up the database.

\bibliography{references}
\bibliographystyle{unsrt}  

\end{document}